\documentstyle[pre,aps]{revtex}
\begin{document}
\draft
\title{Membrane boundary condition
}
\author{Tadeusz Koszto\l owicz\footnote{Electronic address:
tkoszt@pu.kielce.pl}}

\address{Institute of Physics, Pedagogical University,\\
ul. Konopnickiej 15, PL - 25-406 Kielce, Poland}

\author{Stanis\l aw Mr\' owczy\' nski\footnote{Electronic address:
mrow@fuw.edu.pl}}

\address{ So\l tan Institute for Nuclear Studies,\\
ul. Ho\.za 69, PL - 00-681 Warsaw, Poland \\
and Institute of Physics, Pedagogical University,\\
ul. Konopnickiej 15, PL - 25-406 Kielce, Poland }

\date{10-th March 2000}
 
\maketitle

\begin{abstract}

Using a microscopic phase-space model of the membrane system, the 
boundary condition at a membrane is derived. According to the condition, 
the substance flow across the membrane is proportional to the difference 
of the substance concentrations at the opposite membrane surfaces. The 
Green's function of the diffusion equation is found for the derived boundary 
condition and the exact solution of the equation is given.

\end{abstract}

\vspace{0.5cm}
PACS number(s): 82.65.Fr, 66.10.Cb


\section{Introduction}

The membrane systems play an important role in several fields of 
technology \cite{Rau89}, where the membranes are used as filters, 
and biophysics \cite{Tho76}, where the membrane transport is 
crucial for the cell physiology. The diffusion in the membrane 
system is also interesting by itself as a nontrivial stochastic problem.
The process is usually described by means of the diffusion equation. 
Since the equation is of the second order one needs two boundary 
conditions at the membrane surfaces. One condition is provided by 
the particle number conservation but there is no obvious choice of 
the second condition. Although several proposals can be found in 
the literature \cite{Hoo76,Kos96} a systematic derivation
of the boundary condition within a microscopic model of the process
is missing. 

While the problem of diffusion in a system with a partially permeable
wall (the membrane) is not well explored, there is a vast literature 
devoted to the  systems with (partially) reflecting or (partially) absorbing 
walls. Two classes of microscopic approaches have been repeatedly used 
to derive the boundary condition for these cases: the random walk on the 
discrete lattice, see e.g. \cite{Cha43,Kam72,Wei94}, and the diffusion in the
(continuous) phase-space, see e.g. \cite{Raz82,Cha83}. The first approach
starts with the death-birth equation while the second one usually utilizes 
the Fokker-Planck equation. 

In this note we refer to a simple phase-space model of the membrane system 
and, following \cite{Raz82,Cha83}, we derive, for the first time to our best
knowledge, the boundary condition for the diffusion equation. The condition 
appears to be rather natural - the substance flow across the membrane is 
proportional to the difference of the substance concentrations at the membrane. 
In the second part of our paper we find the Green's function of the diffusion 
equation which satisfy the derived boundary condition. The exact solution of 
the equation with a specific initial condition is also given. Finally, our 
considerations are critically discussed.

\section{Derivation of boundary condition}

Let consider a system where the substance diffuses across the 
membrane which is treated as an infinitely thin partially permeable 
wall located at $x=0$.  Since the concentration gradients are assumed 
to be only along the $x-$direction the problem is effectively
one dimensional. One can imagine the membrane as a reflecting wall 
with homogeneously distributed holes. Then, the substance particles are 
either reflected by the wall or pass through it without change of their 
momenta. We denote by $\alpha$ the ratio of the total surface of all 
holes to the membrane surface. Then, the probability to pass through the 
membrane just equals $\alpha$. The limiting case $\alpha = 0$ 
corresponds to the holeless, not permeable wall while $\alpha = 1$ to 
the fully permeable wall or to lack of membrane at all. The parameter
$\alpha$ is, obviously, independent within our model of the particle 
velocity .

The distribution function of the substance particles $f(p,x,t)$, where $p$ 
and $x$ are, respectively, the particle momentum and position at the time 
$t$, satisfies the relations
\begin{eqnarray}\label{relation1}
f(-p,0^-,t) &=& (1 - \alpha )\; f(p,0^-,t)  + \alpha \; f(-p,0^+,t) \;,
\\[2mm] \label{relation2}
f(p,0^+,t) &=& (1 - \alpha )\; f(-p,0^+,t)  + \alpha \; f(p,0^-,t) \;.
\end{eqnarray}
Defining the particle flow and the partial flows as
\begin{eqnarray} \nonumber
j(x,t) &\buildrel \rm def \over =&
\int_{-\infty}^{+\infty}dp \:{p\over m}\: f(p,x,t)
\\[2mm] \label{flow-plus}
j_+(x,t) &\buildrel \rm def \over =&
\int_0^{+\infty}dp \:{p\over m} \: f(p,x,t)
\\[2mm] \label{flow-minus}
j_-(x,t) &\buildrel \rm def \over =&
-\int_{-\infty}^0 dp \:{p\over m}\: f(p,x,t)
\end{eqnarray} 
with $m$ being the particle mass, one rewrites the relations 
(\ref{relation1},\ref{relation2}) as
\begin{eqnarray}\label{flow-relation1}
j_-(0^-,t) &=& (1 - \alpha )\; j_+(0^-,t)  + \alpha \; j_-(0^+,t) \;,
\\[2mm] \label{flow-relation2}
j_+(0^+,t) &=& (1 - \alpha )\; j_-(0^+,t)  + \alpha \; j_+(0^-,t) \;.
\end{eqnarray}
One observes that adding 
Eqs.~(\ref{flow-relation1},\ref{flow-relation2}) and using the
formula $j(x,t) = j_+(x,t) -j_-(x,t)$ we get the conservation
of the substance flow at the membrane i.e.
\begin{eqnarray}\label{conservation}
j(0^-,t) = j(0^+,t) \;.
\end{eqnarray}

Now, we decompose the distribution function as
\begin{equation}\label{decompose}
f(p,x,t) = f_0(p) + f_1(p,x,t) \;,
\end{equation}
where $f_0(p)$ is equilibrium distribution function i.e.
\begin{eqnarray*}
f_0(p,x,t) = {n(x,t) \over \sqrt{2 \pi mk_BT}}\; 
\exp\Big[{-{p^2 \over 2mk_BT}}\Big]
\end{eqnarray*}
with $T$ and $k_B$ denoting the temperature and Boltzmann constant,
respectively. In contrast to the equilibrium distribution function, which 
is even ($f_0(-p) = f_0(p)$), the function $f_1$ is assumed to be odd 
($f_1(-p,x,t) = - f_1(p,x,t)$). The decomposition (\ref{decompose}),
which plays a crucial role in our considerations, can be justified in two
closely related ways. One can assume that $f(p,x,t)$ satisfies the 
Fokker-Planck equation. Then, the decomposition (\ref{decompose}) 
corresponds to the first two terms of  the expansion of the distribution
function in the large friction limit \cite{Cha83}. One can also refer 
here to a more general scheme of Chapman-Enskog expansion which
is applicable not only for the Fokker-Planck equation. However, the 
function $f_1$ is not obligatory odd in this case. Nevertheless, this
is the odd part of $f_1$ which really matters; the even part does not
contribute to the particle flow while its contribution to the particle 
density can be neglected because $f_0 \gg |f_1|$. At the end of our 
paper we briefly discuss how our results would be modified
if the decomposition (\ref{decompose}) is not limited to the two 
terms.

Substituting the distribution function of the form (\ref{decompose}) 
into Eqs.~(\ref{flow-plus},\ref{flow-minus}) and applying 
the Fick law 
\begin{equation}\label{Fick}
j(x,t) = - D \; {\partial n(x,t) \over \partial x} \;,
\end{equation}
where $D$ is the diffusion constant,
one finds 
\begin{eqnarray}\label{flow-plus-2}
j_+(0^{\pm},t) & = & \sqrt{k_BT \over 2\pi m}\; n(0^{\pm},t)
- {D \over 2} \; {\partial n(x,t) \over \partial x} \Big|_{x=0^{\pm}}
\\[2mm] \label{flow-minus-2}
j_-(0^{\pm},t) & = & \sqrt{k_BT \over 2\pi m}\; n(0^{\pm},t)
+ {D \over 2} \; {\partial n(x,t) \over \partial x} \Big|_{x=0^{\pm}} \;.
\end{eqnarray} 

The relations (\ref{flow-plus-2},\ref{flow-minus-2}) allow one
to convert Eqs.~(\ref{flow-relation1},\ref{flow-relation2}) to
the form
\begin{eqnarray}\label{flow-relation3}
\alpha \sqrt{k_BT \over 2\pi m} \; \Big( n(0^-,t) - n(0^+,t) \Big)
&=& -{2 - \alpha  \over 2}\; D \; 
{\partial n(x,t)\over \partial x} \Big|_{x=0^-}
+{ \alpha  \over 2}\; D \; 
{\partial n(x,t)\over \partial x} \Big|_{x=0^+} \;,
\\[2mm] \label{flow-relation4}
\alpha \sqrt{k_BT \over 2\pi m} \; \Big( n(0^+,t) - n(0^-,t) \Big)
&=& + {2 - \alpha  \over 2}\; D \; 
{\partial n(x,t) \over \partial x} \Big|_{x=0^+}
-{ \alpha  \over 2}\; D \; 
{\partial n(x,t)\over \partial x} \Big|_{x=0^-}\;.
\end{eqnarray}
As seen,  Eqs.~(\ref{flow-relation3},\ref{flow-relation4}) provide
again the flow conservation (\ref{conservation}) and the desired
boundary condition
\begin{equation}\label{boundary}
j(0,t) = - \kappa \;
\Big( n(0^+,t) - n(0^-,t) \Big) \;
\end{equation}
where
$$
j(0,t) = j(0^+,t) = j(0^-,t) \;,
$$
and the membrane permeability coefficients $\kappa$ is defined as
\begin{equation}\label{kappa}
\kappa \equiv {\alpha \over 1 -\alpha} \sqrt{k_BT \over 2\pi m} \;.
\end{equation}

\section{Solving the diffusion equation}

In this section we are going to solve the diffusion equation
\begin{equation}\label{diffusion}
{\partial n(x,t) \over \partial t} = 
D \; {\partial^2 n(x,t) \over \partial x^2} \;,
\end{equation}
with the boundary conditions (\ref{conservation},\ref{boundary}) and the 
initial one given as
$$
n(x,0) = n_0(x) \;.
$$

We introduce, as usually, the Green's function $G(x,x',t,t')$ which solves 
the equation
\begin{equation}\label{G-diffusion}
{\partial \over \partial t} \: G(x,x',t,t')  = 
D \; {\partial^2 \over \partial x^2} \: G(x,x',t,t') \;,
\end{equation}
with the initial condition
\begin{equation}\label{G-initial}
\lim_{t \to t'}{G(x,x',t,t')}  = \delta (x-x') \;,
\end{equation}
and four boundary conditions: two which are obtained from 
(\ref{conservation},\ref{boundary}) i.e. 
\begin{eqnarray}\label{G-boundary1}
  {\partial \over \partial x} \: G(x,x',t,t')\Big|_{x=0^-}
&=& {\partial \over \partial x} \: G(x,x',t,t')\Big|_{x=0^-}
\\ [2mm] \label{G-boundary2}
-D {\partial \over \partial x} \: G(x,x',t,t')\Big|_{x=0^-}
&=& -\kappa \; \big[G(0^+,t;x',t') - G(0^-,t;x',t') \big] \;,
\end{eqnarray}
and
\begin{equation}\label{G-boundary3}
\lim_{x \to \pm \infty}{G(x,x',t,t')}  = 0 \;.
\end{equation}
Having the Green's function the solution of the equation
(\ref{diffusion}) is given as
\begin{equation}\label{solution}
n(x,t) = \int dx'\; n_0(x') \; G(x, x', t,0) \;.
\end{equation}
Further, we always put $t'= 0$ and denote the Green's function
as $G(x,x',t)$.

We find the Green's function applying the standard procedure 
\cite{Car59} of the Laplace transformation which gives
$$
\widetilde G(x,x',s) \buildrel \rm def \over = 
\int _0^{\infty}dt \: e^{-st}\: G(x,x',t) \;.
$$
Transforming Eq.~(\ref{G-diffusion}) and the initial condition 
(\ref{G-initial}) we get the equation
\begin{equation}\label{G-diffusion-L}
D {d^2 \over d x^2} \: \widetilde G(x,x',s)  
- s \: \widetilde G(x,x',s) = - \delta (x-x') \;.
\end{equation}
One trivially finds the general solution of the homogenous equation
(\ref{G-diffusion-L}) while the inhomogenous equation is easily 
solved by means of the Fourier transformation. In this way we get
$$
\widetilde G(x,x',s) = {1 \over 2 D q} \: e^{-q\vert x-x'\vert} 
+ A \: e^{qx} + B \: e^{-qx} \;,
$$
where $q^2 \equiv s/D$ while the constants $A$ and $B$ are determined 
by the boundary conditions 
(\ref{G-boundary1}, \ref{G-boundary2}, \ref{G-boundary3}). To calculate 
the constants one has to distinguish four cases related to the signs of 
$x$ and $x'$. We denote as $G_{++}$ the Green's function which 
corresponds to $x> 0$ and $x'>0$; as $G_{+-}$ that one for the case of 
$x> 0$ and $x'<0$, etc. After inverting the Laplace transform we get
\begin{eqnarray}\label{Green1}
G_{+-}(x,x',t) = G_{-+}(x,x',t) &=& {\kappa \over D} \:
{\rm exp}\Big( {2\kappa ( | x - x'| + 2 \kappa t) \over D} \Big) \;
{\rm erfc}\Big( { | x - x'| + 4 \kappa t \over 2 \sqrt{Dt}} \Big) \;,
\\ [4mm] \label{Green2}
G_{++}(x,x',t) = G_{--}(x,x',t) &=& {1 \over 2 \sqrt{\pi Dt}}\:
\bigg[{\rm exp}\Big( - {(x-x')^2 \over 4Dt} \Big) +
{\rm exp}\Big( - {(x+x')^2 \over 4Dt} \Big) \bigg] 
\\ [2mm] \nonumber
&-&{\kappa \over D} \:
{\rm exp}\Big( {2\kappa ( | x + x'| + 2 \kappa t ) \over D} \Big) \;
{\rm erfc}\Big({| x + x'| + 4 \kappa t \over 2 \sqrt{Dt}}\Big) \;,
\end{eqnarray}
with ${\rm erfc}(x)$ being the complementary error function i.e.
$$
{\rm erfc}(x)= {2 \over \sqrt{\pi}}\int_x^{\infty}dt \, e^{-t^2} \;.
$$

As an application of the Green's functions (\ref{Green1},\ref{Green2})
we consider the time evolution of the concentration of the substance 
which is initially homogeneously distributed in the left half-space
i.e. the initial condition reads
$$
n(x,0) = n_0 \: \Theta (-x) \;.
$$
Then, Eq. (\ref{solution}) provides
\begin{eqnarray}\label{solution-plus}
n_+(x,t) &=& n_0 \int_{-\infty}^0 dx' \; G_{+-}(x,x',t) 
\\ [2mm] \nonumber 
&=& {n_0 \over 2} \; \bigg[
{\rm erfc} \Big({x \over 2 \sqrt{Dt}}\Big) -
{\rm exp}  \Big({2\kappa ( x + 2 \kappa t ) \over D} \Big) \;
{\rm erfc} \Big({x + 4\kappa t \over 2 \sqrt{Dt}}\Big) \bigg]\;,
\\[4mm] \label{solution-minus}
n_-(x,t) &=& n_0 \int_{-\infty}^0 dx' \; G_{--}(x,x',t) 
\\ [2mm] \nonumber 
&=& { n_0 \over 2} \; \bigg[ 2 - 
{\rm erfc} \Big({- x \over 2 \sqrt{Dt}}\Big) +
{\rm exp}  \Big({2\kappa ( - x + 2 \kappa t ) \over D} \Big) \;
{\rm erfc} \Big({- x + 4\kappa t \over 2 \sqrt{Dt}}\Big) \bigg]\;,
\end{eqnarray}
where
$n_+(x,t) \equiv n(x,t)$ for $x > 0$ and 
$n_-(x,t) \equiv n(x,t)$ for $x < 0$.

Let us briefly discuss the solution 
(\ref{solution-plus}, \ref{solution-minus}). One observes that when 
$\kappa = 0$, which corresponds to the fully reflecting wall, we get 
from (\ref{solution-plus}, \ref{solution-minus}) an expected result 
i.e. $n_+(x,t) = 0$ and $n_-(x,t) = n_0$. In the opposite limit 
$\kappa \rightarrow \infty$ corresponding to the lack of any membrane, 
we find the well known formulas:
\begin{eqnarray*}
n_+(x,t) &=&{ n_0 \over 2}\;
{\rm erfc} \Big({x \over 2 \sqrt{Dt}}\Big) \;,
\\[4mm] 
n_-(x,t) &=& {n_0 \over 2}
\bigg[ 2 - {\rm erfc} \Big({- x \over 2 \sqrt{Dt}}\Big) \bigg] \;.
\end{eqnarray*}
Finally, one observes that for $\sqrt{Dt} \gg |x|$ and $\kappa t \gg |x|$ 
the solution (\ref{solution-plus}, \ref{solution-minus}) provides
$$
n_{\pm}(x,t) \cong {n_0 \over 2} 
\Big[ 1 \mp {1 \over 2\kappa}\sqrt{\pi D \over t}\; \Big] \;.
$$
Thus, $n_{\pm}(x,t) = n_0 / 2$ for $t \rightarrow \infty$.

\section{Discussion}

The decomposition (\ref{decompose}), which, as already mentioned,
plays a key role in the boundary condition derivation, includes only
two terms. Since the solution of the Fokker-Planck equation can be
systematically expanded in the powers of the inverse friction 
coefficient, one can easily supplement, following \cite{Cha83}, the 
decomposition (\ref{decompose}) by a next order term. 
Assuming that the Fick law (\ref{Fick}) still holds, we
again get the current conservation (\ref{conservation}) while the
equation analogous to (\ref{boundary}) reads:
\begin{equation}\label{boundary2}
j(0,t) + {\alpha  \over 1 - \alpha }\; {D \over 4 \sqrt{\pi} \gamma} \;
\bigg( {\partial^2 n(x,t)\over \partial x^2} \Big|_{x=0^+}
- {\partial^2 n(x,t)\over \partial x^2} \Big|_{x=0^-} \bigg)
= - \kappa \; \Big( n(0^+,t) - n(0^-,t) \Big) \;,
\end{equation}
where $\gamma $ is the friction coefficient from the Fokker-Planck
equation. If, following \cite{Raz82}, one treats LHS of Eq.~(\ref{boundary2})  
as first two terms of  the Taylor expansion of the current at a finite $x$, 
the boundary condition (\ref{boundary2}) can be manipulated to the form
\begin{equation}\label{boundary3}
(1 - k) \: j\Big(-{x_0 \over 1 -k},t\Big) 
+ k \: j\Big({x_0 \over k},t \Big) 
= - \kappa \; \Big( n(0^+,t) - n(0^-,t) \Big) \;,
\end{equation}
where $k$ is an arbitrary number while $x_0$ denotes the thickness 
of the kinetic near membrane layer equal
$$
x_0  = { \alpha \over 4 \sqrt{\pi} \: (1 - \alpha ) \: \gamma} \;.
$$
Unfortunately, we can not see a way to remove the arbitrarness of  
$k$ which appears when, due to Eq. (\ref{conservation}), we express 
$j(0,t)$ as $(1-k)j(0^+,t) + kj(0^-,t)$. Such a problem is absent 
in the formula analogous to (\ref{boundary3}) for the case of absorbing 
wall \cite{Raz82}. The results (\ref{boundary2}, \ref{boundary3}) are 
interesting by themselves but they do not seem to be very useful as 
membrane boundary conditions.  As well known, the diffusion equation 
provides a reliable description if the Chapman-Enskog expansion 
converges fast. Then, the third term is not needed in the decomposition 
(\ref{decompose}) and one should use the boundary condition 
(\ref{boundary}). However, the distribution function can significantly
deviate from the equilibrium form in the near membrane layer as it 
happens in the vicinity of the absorbing wall \cite{Bur81}. Then, the
boundary condition (\ref{boundary3}) seems to be natural but the 
diffusion equation should be combined with the Fokker-Plnack for 
a correct description of the kinetic layer and the whole approach
appears to be very cumbersome. 

The formula of the membrane permeability coefficient (\ref{kappa}) 
suggests that the permeability grows proportionally to $\sqrt{T}$. 
However, one should remember that this dependence has been obtained 
within a highly simplified mechanical model of the membrane which is 
treated as an infinitely thin reflecting wall with homogeneously 
distributed holes. Then, the substance particles are either reflected 
by the wall or pass through it without change of their momenta. In 
reality, the membrane structure is much more complicated and the 
parameter $\alpha$ should be treated as an effective probability to 
pass through the membrane. Consequently, $\alpha$ might be temperature 
dependent and then $\kappa$ is no longer proportional to $\sqrt{T}$.

In the series of papers of one of us \cite{Kos96}, the boundary condition 
different than (\ref{boundary}) has been advocated. Namely, using the
symmetry arguments of the Green's function, there has been found instead 
of (\ref{boundary}) the relation:
\begin{equation}\label{boundary-kos}
j(0,t) = (1 - \delta) j_0(0,t) \;,
\end{equation}
where $\delta$ is a dimensionless membrane permeability coefficient
and $j_0$ is the substance flow in the system with the removed 
membrane. As discussed in \cite{Kos96}, there is a transparent
probabilistic interpretation of the relation (\ref{boundary-kos}). 
The solution of the diffusion equation, which satisfies the condition 
(\ref{boundary-kos}) \cite{Kos96}, is qualitatively different than that 
found here. According to the solution, there is a finite flow across 
the membrane at infinite time while 
Eqs.~(\ref{solution-plus}, \ref{solution-minus}) give the vanishing 
flow in this limit. 

We hope that an experiment will help to chose a right boundary 
condition and we plan to perform such an analysis in collaboration
with experimentalists. However, one should remember that there is 
a whole variety of the mechanisms of the substance transport 
across the membrane \cite{Rau89}. If the boundary condition 
depends on the actual mechanism the problem does not have 
a unique solution.

\begin{acknowledgments}

We are very grateful to Konrad Bajer for fruitful discussions and
calling our attention to the boundary condition (\ref{boundary}).
This work was partially supported by the Polish Committee of Scientific 
Research under Grant No. 2 P03B 129 16.

\end{acknowledgments}


\end{document}